\documentclass{MyStyle}
\usepackage[square]{natbib}
\usepackage[notablist,nofiglist]{endfloat}
\usepackage{doublespace}

\author{Christian Paroissin\addressmark{1}}
\title[A new graphical tool of outliers detection]{A new graphical tool of outliers detection in regression models based on recursive estimation}

\address{\addressmark{1} Universit\'e de Pau et des Pays de l'Adour, Laboratoire de Math\'ematiques Appliqu\'ees - UMR CNRS 5142, Avenue de l'Universit\'e, 64013 Pau cedex, France, {\tt cparoiss@univ-pau.fr}, Phone number: +33 5 59 40 75 69}



\LastVersion{\today}

\begin{document}
\maketitle

\begin{abstract}
We present in this paper a new tool for outliers detection in the context of multiple regression models. This graphical tool is based on recursive estimation of the parameters. Simulations were carried out to illustrate the performance of this graphical procedure. As a conclusion, this tool is applied to real data containing outliers according to the classical available tools.
\end{abstract}

\section{Introduction}

We consider the classical multiple regression model (or linear model). Let $Y$ be a random vector (called response variable) in $\RR^n$ such that $\EE[Y]=X\beta$ and ${\mbox{cov}}(Y)=\sigma^2 I_n$ where $X \in {\cal{M}}_{n,p}(\RR)$ is a known matrix (rows of $X$ are the explanatory variables) and where $\beta \in  \RR^p$ and $\sigma^2 \in \RR_+$ are the unknown parameters (to be estimated). If the rank of $X$ equals to $p$ (which will be assumed here), then the solution of least-square problem $\widehat{\beta}$ is unique and is given by $\widehat{\beta} = ({}^tXX)^{-1}{}^tXY$. This estimator is unbiased with covariance matrix equal to $\sigma^2({}^tXX)^{-1}$. It follows that the prediction $\widehat{Y}$ is a linear transformation of the response variable $Y$: $\widehat{Y}= HY$ with $H=X({}^tXX)^{-1}{}^tX$ (called hat matrix).
\\[1ex]
Sensitive analysis is a crucial, but not obvious, task. Three important notions can be considered together: outliers, leverage points and influential points. The notion of outlier is not easy to define. In fact one has to distinguish between two cases: an observation can be an outlier with respect to the response variable or/and to the explanatory variable(s). An observation is said to be an outlier w.r.t. the response variable if its residual (standardized or not) is large enough. This notion is not sufficient in some cases as for the fourth Anscombe data set \cite{Anscombe}: the residual of the extreme point is zero but it is clearly an outlier. It follows the second definition of an outlier: an observation is an outlier w.r.t. the explanatory variable(s) if it has a high leverage. As precised by Chatterjee and Price \cite{ChatterjeePrice},''the leverage of a point is a measure of its 'outlyingness' [\ldots] in the [explanatory] variables and indicates how much that individual point influences its own prediction value''. A classical way to measure leverage is to consider diagonal elements of the hat matrix $H$ (that depends only on matrix $X$ and not on $Y$): the $i$-th observation is said to have a high leverage if $H_{ii} \geqslant 2p/n$ (which is twice the average value of the diagonal elements of $H$). Any observation with a high leverage has to be considered with care. From the above quotation, one has also to define the notion of influential observations. An observation is ''an influential point if its deletion, singly or in combination with others [...] causes substantial changes in the fitted'' \cite{ChatterjeePrice}. There exists several measures of influence: among the most widely used, the Cook distance \cite{Cook} and the DFFIT distance \cite{BelsleyKuhWelsch}. These two distances are cross-validation (or jackknife) methods since they are defined on regression with deletion of the $i$-th observation (when measuring its influence). Observations which are influential points have also to be considered with care. However one has to consider simultaneously leverage and influence measures. Cook procedure has been improved and used for defining several procedures (see \cite{PenaYohai} for instance). For a survey about various methods for multiple outliers detection throughout Monte Carlo simulations, the reader could refer to \cite{WisnowskiMontgomerySimpson}.
\\[1ex]
In this paper we propose a new graphical tool for outliers detection in linear regression models (but not for the identification of the outlying observation(s)). This graphical method is based on recursive estimation of the parameters. Recursive estimation over a sample provides a useful framework for outliers detection in various statistical models (multivariate data, time series, regression analysis, \ldots). Next section is devoted to the introduction of this tool.  In order to study its performance, simulations were carried out on which our tool was applied in section 3. First we apply our graphical method to the case of data set with one single outlier, either in the explanatory variable or/and in the response variable. Second we apply the graphical tool to the case of multiple outliers. In the last section, our tool is applied to real data for which it is well-known that they contain one or two outliers.

\section{A new graphical tool}

In one hand many authors suggested graphical tools for the outliers detection in regression models. For instance Atkinson \cite{Atkinson-81} suggested half normal plots for the detection of single outlier (see also \cite{Atkinson-book} for a large panorama). In other hand the seminal paper by Brown {\em et al.} \cite{BrownDurbinEvans} (see also \cite{PhillipsHarvey}) about recursive residuals (we share Nelder opinion - see his comments about \cite{BrownDurbinEvans} - about the misuse of 'recursive residuals' instead of 'sequential residuals' for instance, but as it is noticed by Brown {\em et al.} \cite{BrownDurbinEvans} ''the usage [of this term] is too well-establish to change'') has been the source of various studies on outliers or related problems, most of them being based on CUSUM test. Schweder \cite{Schweder} introduced a related version of CUSUM test, the backward CUSUM test (the summation is made from $n$ to $i$ with $i \geqslant p+1$) which was proved to have greater average power (than the classical CUSUM test). Later Chu {\em et al.} \cite{ChuHornikKuan} proposed MOSUM tests based on moving sums of recursive residuals.
\\[1ex]
Comments by Barnett and Lewis \cite{BarnettLewis} about recursive residuals summarize well all the difficulty when considering such approach: ''There is a major difficulty in that the labeling of the observations is usually done at random, or in relation to some concomitant variable, rather than 'adaptively' in response to the observed sample values". For instance Schweder \cite{Schweder} in order to develop two methods of outlier detection assumed that the data set could be divided into two subsets with one containing no outliers. In \cite{HadiSimonoff} the reader will find another case in which the half sample is used and assumed to be free of outliers. Since these methods are not satisfactory Kianifard and Swallow \cite{KianifardSwallow-89} defined a test procedure for the outliers detection applied to data ordered according to a given diagnostic measure (standardized residuals, Cook distance, \ldots). Notice that recursive residuals can also be used to check the model assumptions of normality and homoscedasticity \cite{GalpinHawkins,HedayatRobson}. For a review about the use of recursive residuals in linear models, the reader could refer to the state-of-art in 1996 by Kianifard ans Swallow \cite{KianifardSwallow-96} (see also a less recent state-of-art by Hawkins \cite{Hawkins}).
\\[1ex]
For a given subset of observations, estimators of the parameters are invariant under any permutation of the observations, except if one apply recursive estimations. The idea of a (graphical or not) method based on recursive estimation (of the parameters) is to order the observations such that the presence of one or more outliers will be visible (on a figure or/and on a table). This point of view was used by Kianifard and Swallow \cite{KianifardSwallow-89} in the method described above. However their procedure does not guarantee that outliers are detected: this unfortunate case happens for instance when the outliers is precisely one of the $p$ first observations (which are used for the initialization of the recursive computation of residuals). This point has been already noticed by Clarke \cite{Clarke} (who focused on robust method of outlier detection in the case of small sample size). For example, one can clearly observe this phenomenon on the fourth Anscombe data set \cite{Anscombe} if one uses the standardized residuals as a diagnostic measure (see the introduction above for previous comments on this data set). For each of these cases it is usually assumed that the initial subset (or elemental set) does not contain outliers  (such subset are called to be clean subset) but with no guarantee that this assumptions is checked (see \cite{HawkinsBraduKass} for another such situation).
\\[1ex]
Since the graphical tool we propose is based on a recursive procedure, we will introduce some notation for parameters estimation based on a subset of the observations. For any subset $I$ of $\{1, \ldots, n\}$, we denote by $\widehat{\beta}(I)$ the estimator of $\beta$ based on observations $X_{i1}, \ldots, X_{ip}$ with $i \in I$. We denote by $X_I$ (resp. $Y_I$) the sub-matrix of $X$ (resp. $Y$) corresponding to the above situation. We will assume that for any subset $I$ such that $|I| \geqslant p$ the matrix $X_I$ is full-rank. It follows that $\widehat{\beta}(I)$ is unique and given by $\widehat{\beta}(I) = ({}^tX_IX_I)^{-1}{}^tX_IY_I$. We will denote by $S_n$ the set of all permutations of $\{1, \ldots, n\}$ and for any permutation $s \in S_n$, $I_i^s := \{s(1), \ldots, s(i)\}$.
\\[1ex]
The graphical procedure we suggest here consists in generating $p$ different graphical displays, one for each coordinates of $\beta$ (including the intercept in case of). On the $j$-th graphical display points $(i,\widehat{\beta}_j(I_{p+i-1}^s))$ with $i \in \{1, \ldots, n-p+1\}$ are plotted, for a given number of permutations $s \in S_n$ (points can be joined with lines).  Similar graphical displays can also be produced for the variance estimation and for various coefficients (determination coefficient, AIC, \ldots). This graphical tool can be viewed as dynamic graphics defined by Cook and Weisberg \cite{CookWeisberg-89}. This approach seems to be new to the best of our knowledge despite recursive residuals are quite old (indeed earlier related papers are due to Gauss in 1821 and Pizzetti in 1891 - see the historical note by Farebrother \cite{farebrother} ; see also \cite{Plackett}). In fact recursive residuals and recursive estimation are most of the times considered in the context of time series (see for instance the presentation proposed in \cite{BelsleyKuhWelsch}) since hence there exists a natural order for the observations. It follows that in such situation it is not possible to consider any permutation of the observations (it explains why recursive residuals are mainly used to check the constancy of the parameters over time).
\\
The presence of one (or more) outlier in a data set should induce jumps/perturbations at least on some of these plots. However the effect will not be really visible if the outlier lies in the first observations (see above the remark above about \cite{HawkinsBraduKass} and \cite{KianifardSwallow-89}) or in the last observations. In fact, in the first case, the effect will be diluted due to the small sample size inducing a lack of precision in the estimations. And in the second case the effect should be also diluted because of a kind of law of large numbers (as noticed by Anderson in his comments of \cite{BrownDurbinEvans}, $\widehat{\beta}_n$ converges to $\beta$ in probability as $n$ tends to infinity if $({}^tX_nX_n)^{-1}$ converges to zero as $n$ tends to infinity). Hence it suggests that there exists some 'optimal' positions for the outlying individuals in order to be detected by a recursive approach. 
\\[1ex]
The number of permutations used for the graphics should depend on the sample size. We suggest to distinct the three following cases:
\begin{enumerate}
\item Large sample size: one can plot points for all the $n$ circular permutations. In this way, on each graphical displays $n$ lines will be represented.
\item Medium or small sample size: if the sample size is not enough large to apply the above rule, one can choose at random $N$ permutations and to plot the $N$ curves corresponding to recursive estimation. The value of $N$ may depend on $n$: the smallest $n$ is, the largest $N$ has to be.
\item Very small sample size: if $n$ is small enough (say smaller than 10), one can plot all the $n!$ sequences on each graphical displays. Such situation could appear in the context of experimental designs for instance.
\end{enumerate}

A major advantage of this new graphical tool is that it does not require the normality assumption. This assumption is generally required in the former outliers detection procedures (especially when using standardized residuals for instance). Moreover it can be performed on data with few observations.
\\[1ex]
Before applying the graphical method described above to simulated data and real data, we wish to consider some practical aspects:
\begin{enumerate}
\item In order to enlight the presence of outliers (indeed this can reduce the effect induced by the lack of data), one could prefer to plot only points $(i,\widehat{\beta}_j(I_{p+i-1}^s))$ for $i \geqslant \lfloor \alpha n \rfloor$ with $\alpha \in (0,1)$. The value of $\alpha$ may depend on the sample size: for small sample size, the value of $\alpha$ could reach up to $25\%$. This could emphasize the cases where the outliers are in the 'optimal' positions.
\item Since the graphical method suggested here relies on recursive estimation of parameters, one wish to apply updating formula as given by Brown {\em et al.} \cite{BrownDurbinEvans}. However one should avoid to use such formula, especially when dealing with large data set, and prefer to inverse matrices for each points (since computers are more reliable and efficient than in the past). In fact using updating formula may induce cumulative rounding-off errors making the graphical method unuseful (this point was already noticed by Kendall in his comments about the paper by Brown {\em et al.} \cite{BrownDurbinEvans}).
\end{enumerate}

For now we will assume that the response variable $Y$ is a Gaussian random vector. We will see how one can use cumulative sum (CUSUM) of recursive residuals in order to get similar graphical displays revealing the presence (or not) of outliers. This is fully inspired by \cite{BrownDurbinEvans} (see also \cite{GalpinHawkins}). In fact as showed by McGilchrist {\em et al.} \cite{McgilchristLiantoBryon} (in a more general context), recursive residuals and recursive estimations of $\beta$ are related one to the other by re-writing the update formula as follows for $i \in \{ 1, \ldots, n-p\}$,
\begin{equation*}
\widehat{\beta}(I^s_{p+i}) = \widehat{\beta}(I^s_{p+i-1}) + \frac{R(I^s_{p+i}) ({}^tX_{I^s_{p+i-1}}X_{I^s_{p+i-1}})^{-1}}{\sqrt{1+{}^tx_{s(p+i)}({}^tX_{I^s_{p+i-1}}X_{I^s_{p+i-1}})^{-1}x_{s(p+i)}}} \;,
\end{equation*}
where $x_i$ denotes the $i$-th row of $X$ and where $R(I^s_{i+1})$ is the $i$-th recursive residuals defined by:
\begin{equation*}
R(I^s_{p+i}) = \frac{Y_{s(p+i)}-{}^tx_{s(p+i)}\widehat{\beta}(I^s_{p+i-1})}{\sqrt{1+{}^tx_{s(p+i)}({}^tX_{I^s_{p+i-1}}X_{I^s_{p+i-1}})^{-1}x_{s(p+i)}}} \;.
\end{equation*}
As proved by Brown {\em et al.} (lemma~1 in \cite{BrownDurbinEvans}), $R(I^s_{1}), \ldots, R(I^s_{n-p})$ are iid random variables having the Gaussian distribution with mean 0 and variance $\sigma^2$. It allows to construct a continuous-time stochastic process using Donsker theorem (see chapter~2 in \cite{Billingsley}):
\begin{equation*}
\forall t \in (0,1) \;, \quad X_n(t) = \frac{1}{\sigma\sqrt{n}} \left( S_{\lfloor nt \rfloor} + (nt-\lfloor nt \rfloor)R(I^s_{\lfloor nt \rfloor+p+1}) \right) \;,
\end{equation*}
where $S_0=0$ and $S_i = S_{i-1} + R(I^s_{p+i})$. The unknown variance $\sigma^2$ is estimated considering all the observations: $\widehat{\sigma}^2 = ||Y-\widehat{Y}||^2/(n-p)$. If all the assumptions of the Gaussian linear model are satisfied, $\{  X_n(t) \,;\,t \in (0,1) \}$ converges in distribution to the Brownian motion as $n$ tends to infinity. It follows that this graphical method could be only used for large sample size. According to Brown {\em et al.} \cite{BrownDurbinEvans}, the probability that a sample path $W_t$ crosses one of the two following curves:
\begin{equation*}
y = 3a\sqrt{t} \quad {\mbox{or}} \quad y = -3a\sqrt{t}
\end{equation*}
equals to $\alpha$ if $a$ is solution of the equation:
\begin{equation*}
1-\Phi(3a) + \exp(-4a^2)\Phi(a) = \frac{1}{2}\alpha
\end{equation*}
where $\Phi$ is the cumulative distribution function of the standard Gaussian distribution (for instance, when $\alpha=0.01$ it gives $a=1.143$).

\section{Simulations}

In this section we provide some simulations in order to observe the phenomenon which arises in such graphical displays in presence of one or more outliers. We will first consider the case where the data set contains only one outlier (either in the explanatory variable or/and in the response variable). Secondly we will consider the case of multiple outliers which is more difficult to detect when using the classical tools.

\subsection{Single outlier}

\begin{figure}
\begin{center}
\includegraphics[width=15cm]{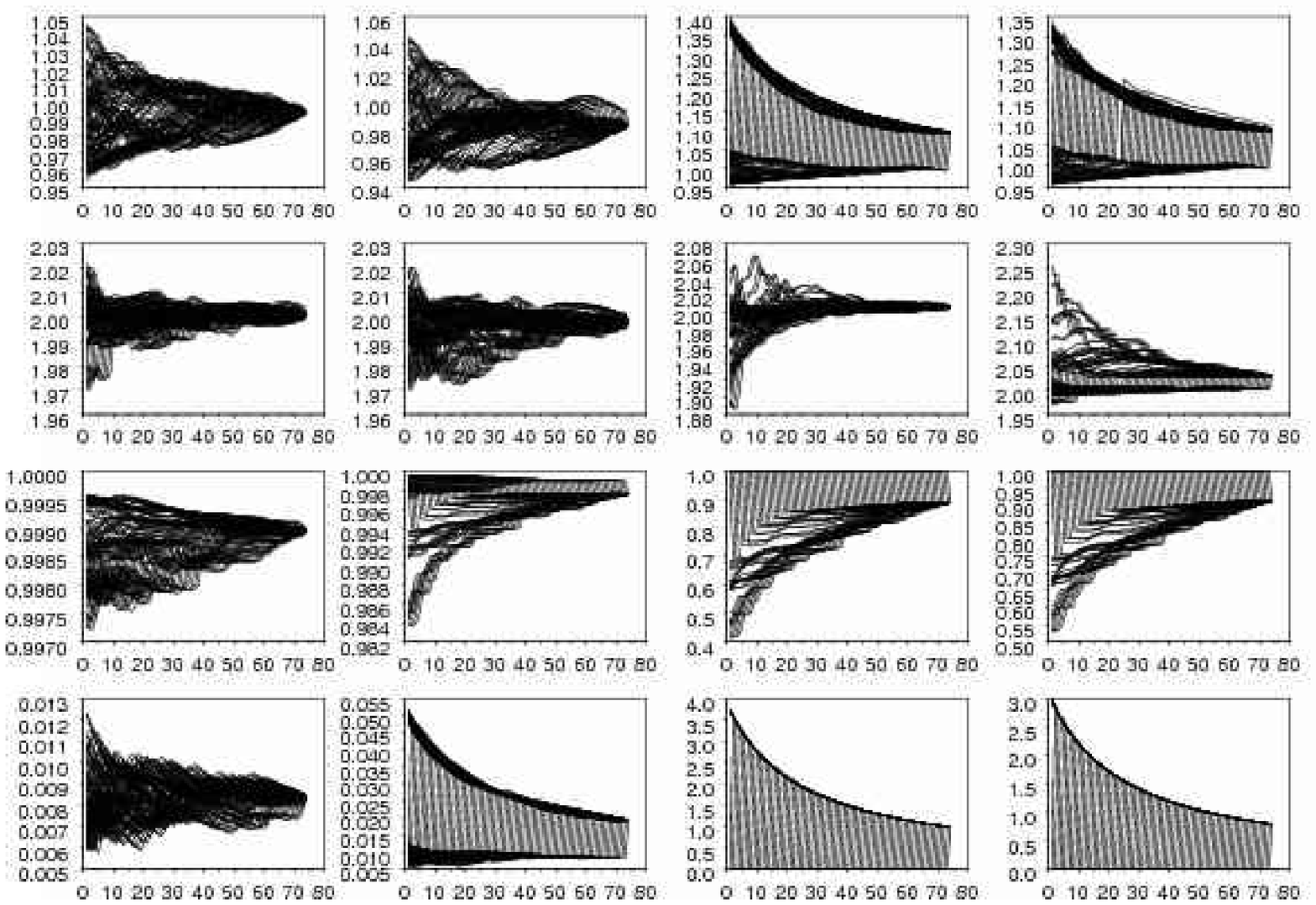}
\caption{Graphical plots for simulated data: single outliers and large sample size}
\label{fig:large-single}
\end{center}
\end{figure}

We present here some simulations on which we apply our graphical tool. Data were generated as follows :
\begin{equation*}
\forall i \in \{ 1, \ldots, n\} \;, \quad y_i = 1 + 2 x_i + \varepsilon_i \;,
\end{equation*}
where $(x_i)$ are iid random variables double exponential distribution with mean 1 and $(\varepsilon_i)$ are iid random variables with the centered Gaussian distribution with standard deviation $\sigma=0.1$. From this model, we derive three perturbed bivariate data sets. First we construct the univariate data set $(\tilde{x}_i)$ as follows: for all $i \in \{1, \ldots, n\} \setminus \{\lfloor n/2 \rfloor\}$ and $\tilde{x}_{\lfloor n/2 \rfloor} = 10 x_{\lfloor n/2 \rfloor}$ (it corresponds to a typo errors with the decimal separator symbol). We construct similarly the perturbed univariate data set $(\tilde{y}_i)$. Thus we combine these univariate data sets to produce four different scenario: no outlier, one outlier in the explanatory variable ($x$), one outlier in the response variable ($y$) and one outlier simultaneously in the explanatory and response variables. 
\\[1ex]
Figure~\ref{fig:large-single} shows these four situations (one for each column) with $n=100$ observations (large sample size): the two first rows contain the recursive estimations of $\beta_0$ and $\beta_1$, the third one the recursive values of $R^2$ (determination coefficient) and the last one the recursive estimations of $\sigma^2$. The presence of one outlier (either in the explanatory variable and/or in the response variable) leads to perturbations in the recursive parameter estimations (especially for the variance estimation) and in the recursive computation of the determination coefficient.

\begin{figure}
\begin{center}
\includegraphics[width=15cm]{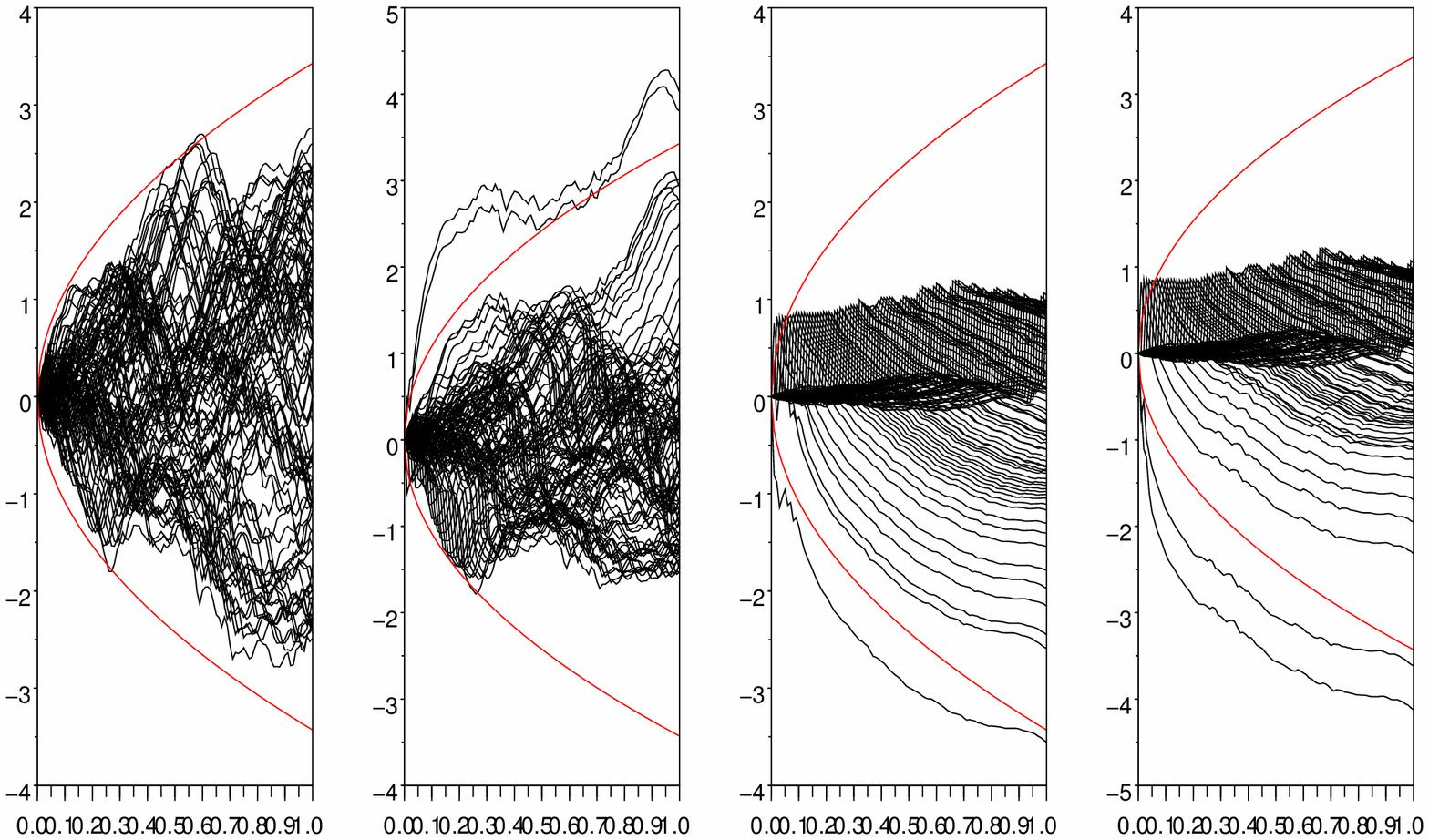}
\caption{CUSUM plots for simulated data: single outliers and large sample size}
\label{fig:large-single-cusum}
\end{center}
\end{figure}

On figure~\ref{fig:large-single-cusum} (each column concern each situation as described above), stochastic processes (that should be Brownian motions in model assumptions are all satisfied) constructed with the CUSUM procedure (see last part in the previous section) are plotted for all circular permutations. Even though these stochastic processes do not frequently cross the parabolic border, it is clear qualitatively of the outliers presence in the three last cases.

\begin{figure}
\begin{center}
\includegraphics[width=15cm]{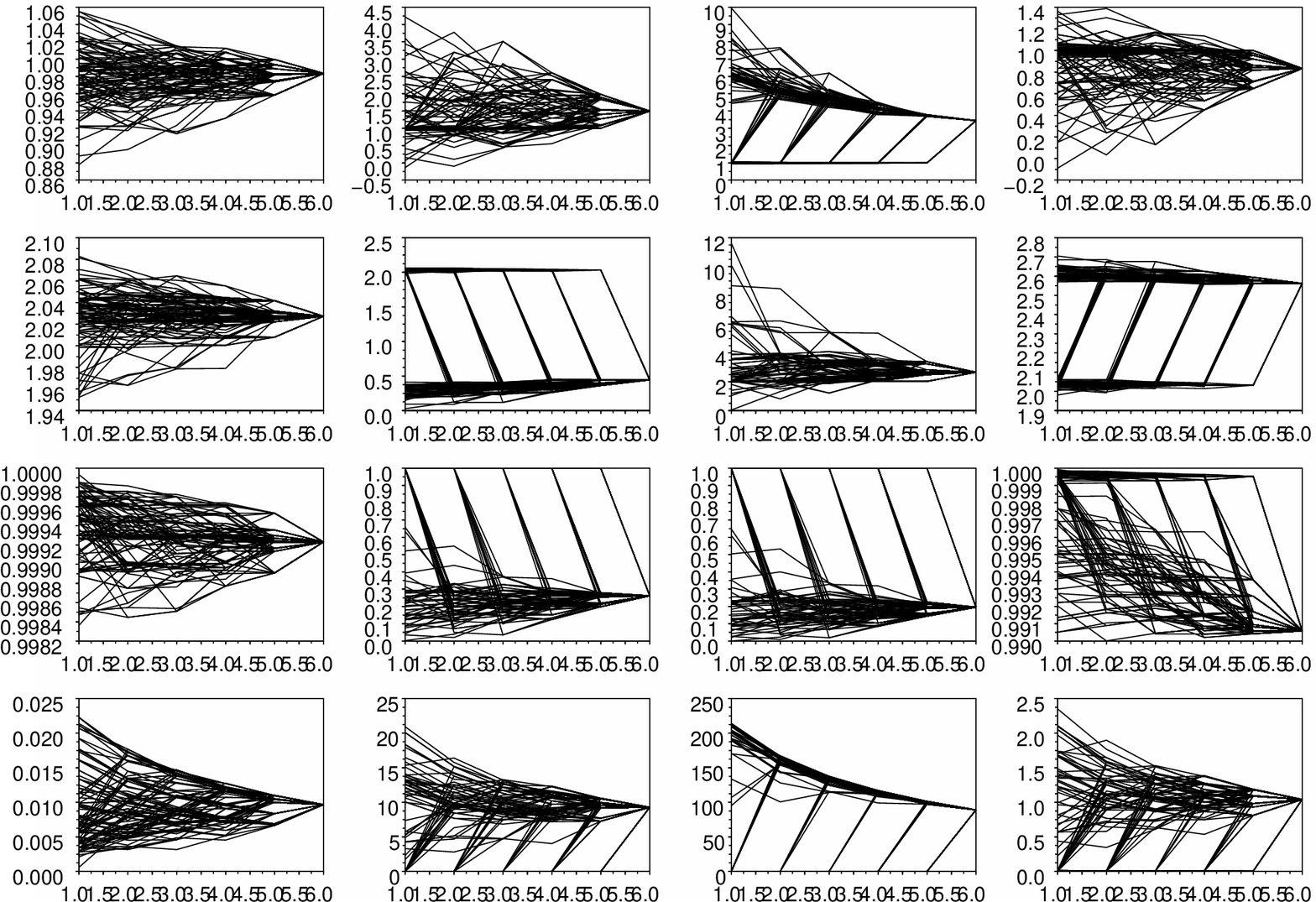}
\caption{Graphical plots for simulated data: single outliers and small sample size}
\label{fig:small-single}
\end{center}
\end{figure}

Figure~\ref{fig:small-single} contains the same outputs (as on the first one) but with $n=10$ (small sample size) and with $N=100$ (the number of random permutations on which recursive estimations are done). Similar outputs are obtained in this case, with slightly difference due precisely to the sample size. The presence of the outlier is more visible for the recursive estimations of $\beta_1$ and for the recursive estimations of the variance $\sigma^2$.

When the response variable $Y$ is a non-Gaussian random vector, the method is still valid an it leads also to the same kind of phenomenon on the various plots. Moreover such approach can be also used to detect switching regime in a regression model. Simulations for these cases were carried out (but not presented here).

\subsection{Multiple outliers}

\begin{figure}
\begin{center}
\includegraphics[width=15cm]{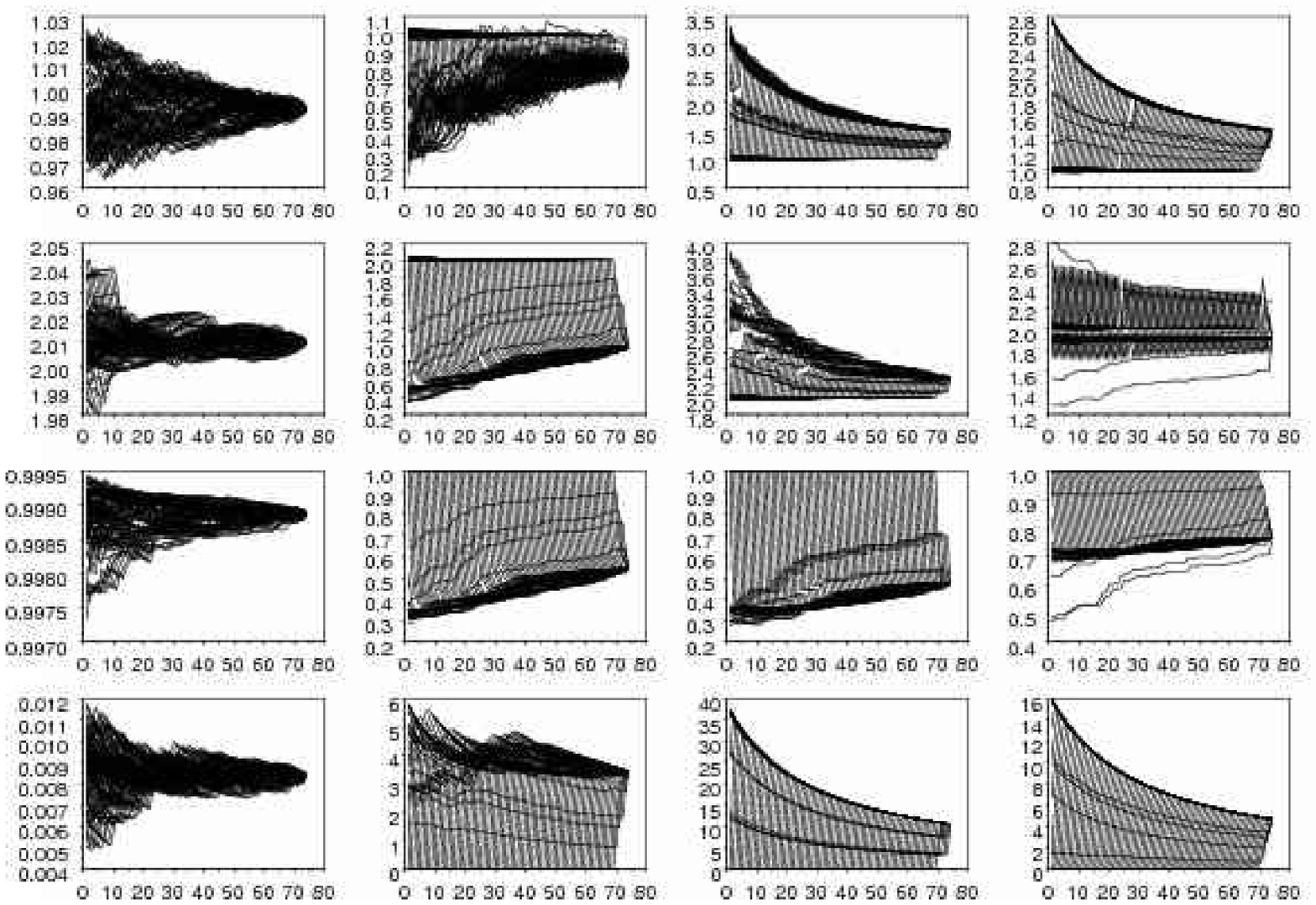}
\caption{Graphical plots for simulated data: multiple consecutive outliers}
\label{fig:large-mult-1}
\end{center}
\end{figure}

\begin{figure}
\begin{center}
\includegraphics[width=15cm]{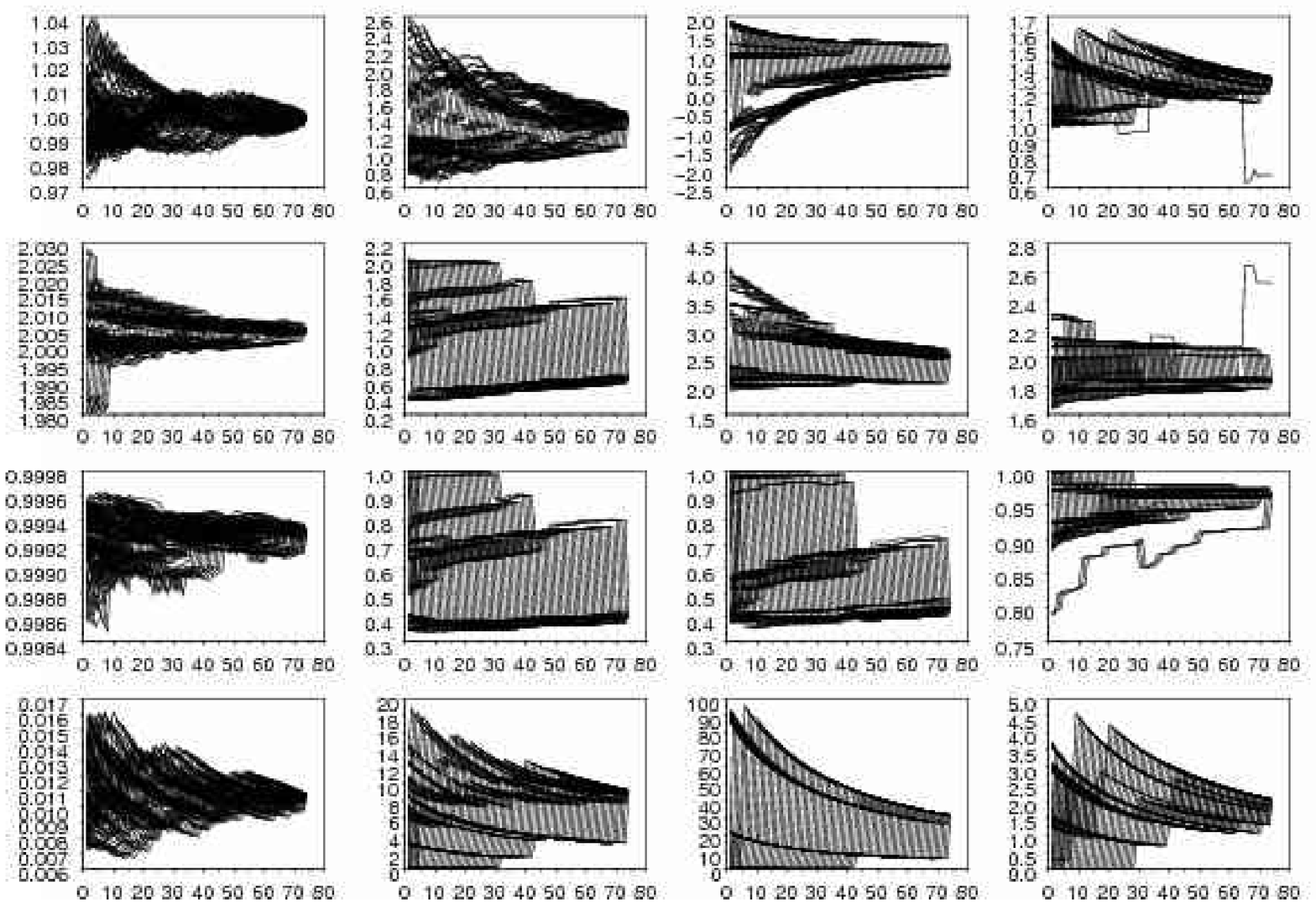}
\caption{Graphical plots for simulated data: multiple outliers randomly chosen}
\label{fig:large-mult-2}
\end{center}
\end{figure}

The presence of multiple outliers in a data set is more difficult to detect. Methods based on single deletion \cite{Atkinson-book,CookWeisberg-82} may fail and thus outliers will be remained undetected. This phenomenon is called the 'masking effect': in presence of multiple outliers, ''least squares estimation of the parameters may lead to small residuals for the outlying observations'' \cite{AtkinsonRiani} (see also \cite{Lawrence} for a discussion about this effect). Moreover  ''if a data set contains more than one outlier, because of the masking effect, the very first observation [with the largest standardized residuals] may not be declared discordant [i.e. as an outlier]'' \cite{Paul}. However since we initialize the recursive estimations at various positions in the data set, this consequence of the masking effect should disappear.
\\[2ex]
We consider the same model as the previous section but in the perturbed univariate data sets we introduce multiple outliers. Two cases are considered: first the outliers are consecutive observations and second the outliers are at random positions in the data sets. Simulation were only carried out for large samples. Figures~\ref{fig:large-mult-1} and~\ref{fig:large-mult-2} contain the outputs obtained respectively with $5$ consecutive outliers and with $5$ outliers uniformly drawn at random over $\{1, \ldots, 100\}$.

\section{Application to health data sets}

We apply our graphical tool to two real data sets. A simple regression will be performed on the first data set which contains a single of outlier. While a multiple regression will be performed on the second data sets which contains a couple of outliers.

\begin{itemize}

\item Alcohol and tobacco spending in Great Britain \cite{MooreMcCabe}. Data comes from a British government survey of household spending in the eleven regions of Great Britain. One can consider the simple regression of alcohol spending on tobacco spending. It appears that this data set contains one single outlier (corresponding to Northern Ireland - the last individual in the data set). On figure~\ref{fig:AT} the various recursive estimations are plotted: from left to right and from up to down, $\beta_0$, $\beta_1$, $R^2$ and $\sigma^2$. Red lines (resp. black) correspond to data with (resp. without) the single outlier. These outputs were obtained by applying the rule for small data sets (with $N=100$ randomly chosen permutations). Graphical plots of the variance estimation and of the determination coefficient clearly indicates the presence of an outlier.

\begin{figure}
\begin{center}
\includegraphics[width=15cm]{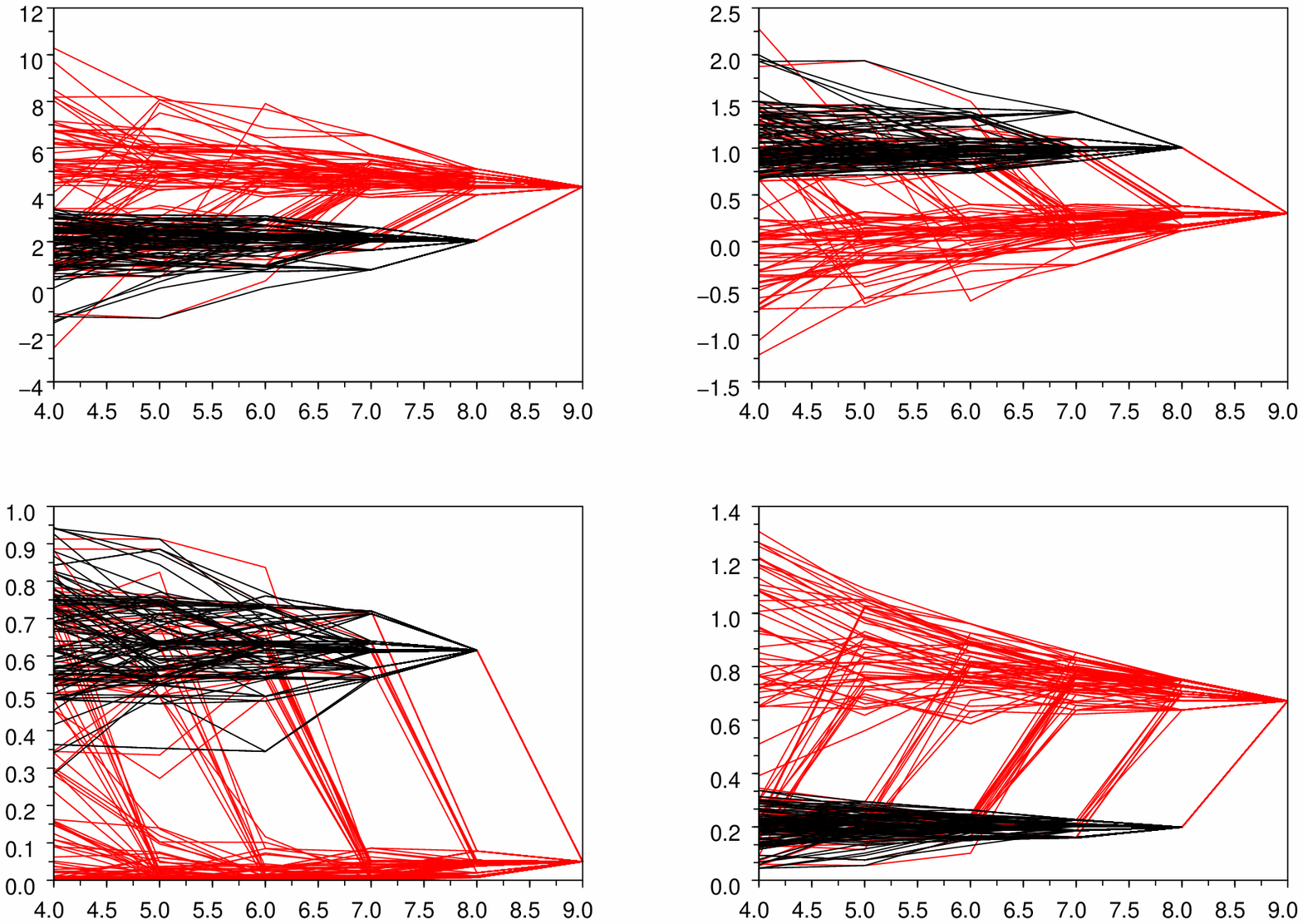}
\caption{Graphical plots for alcohol and tobacco data}
\label{fig:AT}
\end{center}
\end{figure}

\item Smoking and cancer data \cite{Fraumeni}. The data are per capita numbers of cigarettes smoked (sold) by 43 states and the District of Columbia in 1960 together with death rates per thousand population from various forms of cancer: bladder cancer, lung cancer, kidney cancer and leukemia. A classical sensitive analysis leads to conclude that the data set contains two outliers, Nevada and the District of Columbia (the two last individuals in the data set), in the distribution of cigarette consumption (the response variable). Figure~\ref{fig:SC} contains the outputs in three cases (corresponding to the three columns): one of the two outliers have been removed for the two first cases and the two outliers have been removed in the last case. As for the previous example, the red line correspond to the original data set and the red one to the data set with one or two outliers removed. The five first rows contain plots for $\widehat{\beta}$, the sixth row the plot for the determination coefficient and the last row the plot for $\widehat{\sigma}$. The graphical plots for the variance estimation indicates clearly that removing only one outlier is not sufficient. 

\begin{figure}
\begin{center}
\includegraphics[width=15cm]{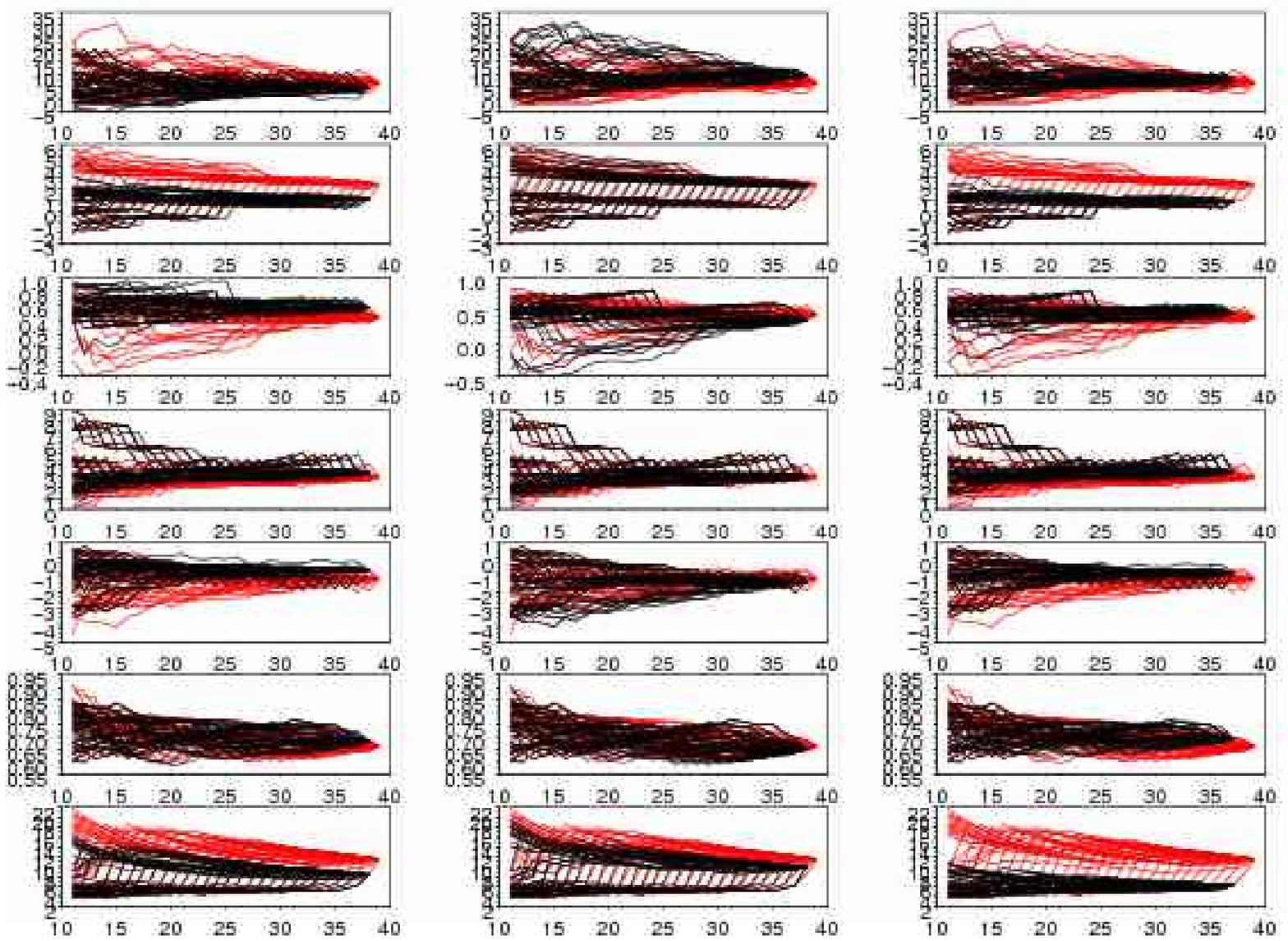}
\caption{Graphical plots for smoking and cancer data}
\label{fig:SC}
\end{center}
\end{figure}

\end{itemize}


\bibliographystyle{plain}

\end{document}